# Detailed experimental and numerical analysis of a cylindrical cup deep drawing: pros and cons of using solid-shell elements

J. Coër[a], H. Laurent[a], M.C. Oliveira[b,1], P.-Y. Manach[a], L.F. Menezes[b]

[a]Univ. Bretagne Sud, FRE CNRS 3744, IRDL, F-56100 Lorient, France

[b]CEMMPRE, Department of Mechanical Engineering, University of Coimbra, Polo II, Rua Luís Reis Santos, Pinhal de Marrocos, 3030-788 Coimbra, Portugal

**Abstract**

The Swift test was originally proposed as a formability test to reproduce the conditions observed in deep drawing operations. This test consists on forming a cylindrical cup from a circular blank, using a flat bottom cylindrical punch and has been extensively studied using both analytical and numerical methods. This test can also be combined with the Demeri test, which consists in cutting a ring from the wall of a cylindrical cup, in order to open it afterwards to measure the springback. This combination allows their use as benchmark test, in order to improve the knowledge concerning the numerical simulation models, through the comparison between experimental and numerical results.

The focus of this study is the experimental and numerical analyses of the Swift cup test, followed by the Demeri test, performed with an AA5754-O alloy at room temperature. In this context, a detailed analysis of the punch force evolution, the thickness evolution along the cup wall, the earing profile, the strain paths and their evolution and the ring opening is performed. The numerical simulation is performed using the finite element code ABAQUS, with solid and solid-shell elements, in order to compare the computational efficiency of these type of elements. The results show that the solid-shell

---

[1]Corresponding author: Tel.: +351239790700, Fax: +351239790701.
E-mail addresses: jeremy.coer@univ-ubs.fr (J. Coër), herve.laurent@univ-ubs.fr (H. Laurent), marta.oliveira@dem.uc.pt (M.C. Oliveira), pierre-yves.manach@univ-ubs.fr (P.-Y. Manach), luis.menezes@dem.uc.pt (L.F. Menezes)

element is more cost-effective than the solid, presenting global accurate predictions, excepted for the thinning zones. Both the von Mises and the Hill48 yield criteria predict the strain distributions in the final cup quite accurately. However, improved knowledge concerning the stress states is still required, because the Hill48 criterion showed difficulties in the correct prediction of the springback, whatever the type of finite element adopted.

**Keywords:** Deep drawing, Springback, Finite element analysis, AA5754-O aluminium alloy, Strain paths

# 1 Introduction

Sheet metal forming is a complex process involving geometric, material and boundary conditions nonlinearities, associated with the large strains and contact between the sheet and the tools. Therefore, with the development of the sheet forming technology, several simulating tests were conceived, to try to reproduce at the laboratory scale, the conditions observed in industrial parts. The main focus of these tests was the analysis of the sheet metal formability. Thus, various formability tests, specific for each of the deformation patterns observed have been developed, covering the major ($\varepsilon_1$) *versus* the minor ($\varepsilon_2$) strain space from stretching ($\varepsilon_1$ and $\varepsilon_2 > 0$) to deep drawing ($\varepsilon_1 > 0$; $\varepsilon_2 < 0$) [1].

Among the deep drawing tests, the Swift test has been widely used and is considered as a standard test by the International Deep-Drawing Research Group (IDDRG) [1]. The test consist on forming a cylindrical cup from a circular blank, using a flat bottom cylindrical punch, imposing a negative minor strain state to the sheet plane, associated to a circumferential compression loading, which typically induces the thickening of the material located in the flange. It has been extensively studied and used as a reference test to evaluate the formability limit of the materials, by calculating the maximum Limiting Drawing Ratio (LDR), which corresponds to the proportion between the maximum diameter of the blank, which can be stamped without breaking, and the diameter of the punch. From a theoretical point of view, this ratio cannot exceed the maximum value of 2.72, as demonstrated in [2], with its value varying between 1.8 and 2.4. On the other hand, from an empirical point of view, the ratio is dependent on the properties of the material, particularly the average anisotropy coefficient [3], or the geometry of the tools, particularly the die radius [4, 5]. On the other hand, several authors have shown, via analytical models [5–7], that the LDR is dependent on the friction coefficient or material parameters such as the yield strength, the yield stress, the normal anisotropy coefficient, or the strain rate sensitivity parameter.

With the advent of the numerical simulation analysis many of these simulating tests became also benchmarks, i.e. reference tests to enable the comparison between experimental and numerical results, in order to improve the numerical methods used (type of finite element, contact description algorithms, etc.) and the constitutive models. For instance, the standard Limiting Dome Height (LDH) recommended by the North American Deep Drawing Research Group (NADDRG), was one of the benchmarks proposed under the Numisheet'1996 conference [8]. The other was the S-rail which focuses another important aspect in sheet metal forming operations, the springback phenomenon [9]. In this context, besides the S-rail, many other benchmarks have being proposed, such as the U-rail (Numisheet' 93) [10], the Unconstrained Cylindrical Bending (Numisheet'2002) [11–13], the Draw/bend test geometry [14], or the Demeri test [15, 16]. This last test consists in cutting a ring from the wall of a cylindrical cup, in order to open it afterwards to observe the relaxation of the stresses induced by the forming operation, i.e. the ring opening is an direct measurement of the springback [17]. These tests were used to improve the numerical prediction of springback, particularly by enabling the study of the strong influence of numerical parameters such as the type, order and integration scheme of finite elements as well as the shape and size of the finite element mesh, but also of the constitutive model adopted [11–14, 16, 18].

The deep drawing of cylindrical cups has also been extensively studied (see e.g. [16, 19, 20]) in order to improve the prediction of forming defects, such as wrinkles as was the case in the Numisheet 2002 [19, 21], or the earing effect, for instance in Numisheet 2014 [22]. The prediction of these type of defects, as well as formability, are quite sensitive to the constitutive model adopted, but also to the algorithms adopted to deal with the contact conditions [4, 22]. Therefore, it is widely used to analyse the influence and the accuracy of the yield criterion adopted, on the prediction of the anisotropic behaviour of the materials and to validate the parameters used in the constitutive laws [21, 23], often identified by inverse analysis [24]. Moreover, the forming of a cylindrical cup may also involve operations such as direct redrawing [25] or reverse drawing (Numisheet '99) [26–28], or even ironing [22, 29, 30] used, for example, in the manufacture of beverage cans [31]. The drawing operation makes it possible to shape a cup from a flat sheet, while redrawing or reverse drawing allow the modification of the cup dimensions, by imposing an even more complex strain path. Finally, the ironing operation consists in reducing the thickness of the wall while retaining the internal diameter of the cup. Experimentally, it is observed that the ironing operation tends to reduce the earing effect which occurs during the deep drawing operation, induced by the anisotropic behaviour of the material, while the other two operations tend to accentuate the amplitude of the ears [32]. The numerical simulation of this type of operation requires an accurate description of the double-side contact conditions experienced by the blank sheet (see e.g. [4]). In this context, the conventional hexahedral 8-node element appears as an interesting alternative, as well as the solid-shell elements, since it allows combining the advantages of the conventional shell elements with a more realistic description of the contact. There are several definitions of solid-shell elements, using various numerical methods, in order to avoid buckling and hourglass modes, some of which have been previous used in sheet metal forming simulations [11, 13].

The focus of this study is the experimental and numerical analyses of the Swift cup test, followed by the Demeri test, performed with the AA5754-O alloy at room temperature. Previous studies indicate that, for the assumed conditions, this test involves a drawing and an ironing operation and that the springback prediction is quite sensitive to the numerical parameters and the constitutive model adopted [33]. Thus, a detailed analysis of the following process variables is performed: the punch force evolution, the thickness distribution measured for three directions, the earing profile, the strain paths and their evolution and the ring opening. The following section contains the description of the test conditions and the experimental results. The numerical model adopted in the standard-implicit version of ABAQUS is described in Section 3. Since the aim of the numerical study is to compare the computational efficiency of solid and solid-shell elements, constitutive models available as standard in ABAQUS are selected, in order to assure that the same implementation strategy is adopted in both cases. The comparison between the experimental and numerical results is performed in Section 4, highlighting the influence of the element type and the constitutive model adopted, in each of the process parameters under analysis. Finally, the main conclusions are presented in Section 5.

## 2   Experimental procedure

This section contains the detailed description of the test conditions considered for the forming process and the posterior springback evaluation using the split-ring test. It should be mentioned that, globally, the test conditions are identical to the ones adopted in the cylindrical cup proposed as benchmark at conference Numisheet 2016 [34]. However, there are same differences, particularly in the die opening radius and the blank thickness that contribute to the change of the process conditions.

### 2.1   Description of the test conditions

The material used in this study was sampled from a rolled sheet of 1-mm gauge AA5754-O aluminium alloy (Al–3%Mg), commonly used in the automotive industry to produce inner body panels. The mechanical behaviour of this material was studied by performing uniaxial tensile tests with the specimen oriented along the rolling direction (RD), 45º to the RD (DD) and the transverse direction to RD (TD), and monotonic shear tests at RD. The results of these tests have been described in detail in [35–37]. In order to help the analysis of the results, the *r*-values determined were: $r_0 = 0.663$; $r_{45} = 0.860$ and $r_{90} = 0.717$. The yield stress is similar for the three directions, with an ultimate tensile strength equal to 222.2 MPa, 211.0 MPa and 216.5 MPa, for the RD, DD and TD, respectively [37].

The deep drawing tests were performed in a sheet metal testing machine (Zwick BUP200), using a tool composed by a die, a blank-holder, a punch and an ejector, which allows extracting the part from the punch at the end of the forming operation. The cutting of the blank is performed automatically using two supplementary tools: the convex cutting blade and the extractor. During this step, the blank holder is used as cutting punch to produce a blank with a diameter of 60 mm ± 0.01. This guarantees the centring of the blank automatically. Figure 1 presents the schematic representation of the tools and their main dimensions are shown in the table in Figure 2. The ratio between the blank and the punch diameter defines a drawing ratio of 1.8. The theoretical value of the gap between the die and the punch is 1.125 mm. However, the measured internal diameter of the die is 35.30 mm, instead of the theoretical value of 35.25 mm, which might be associated with some slight wear of the tool. Therefore, the gap between the die and the punch is 1.15 mm. This means that if the blank with an initial thickness of 1.0 mm thickens more than 0.15 mm, due to the compression stress state in the flange, an ironing stage will occur. The intensity of this ironing stage is strongly dependent of the drawing ratio and of the gap between the die and the punch.

A blank-holder pressure of 1 to 3 MPa should be sufficient to guarantee a part without defects (wrinkles or necking) [38]. However, the machine does not allows applying such low pressure values. An optimal pressure range between 5 and 8 MPa for stamping square cups, made of AA5754-O aluminium alloy is indicated in [39]. Thus, all tests were performed considering a blank-holder force of 6 kN, corresponding to an initial pressure of 4.9 MPa. The tests were performed for a punch velocity of $v_1 = 1.1$ mm.s$^{-1}$, under lubricated conditions (lubricant: Numisheet2002 - Yushiro Form FD-1500).

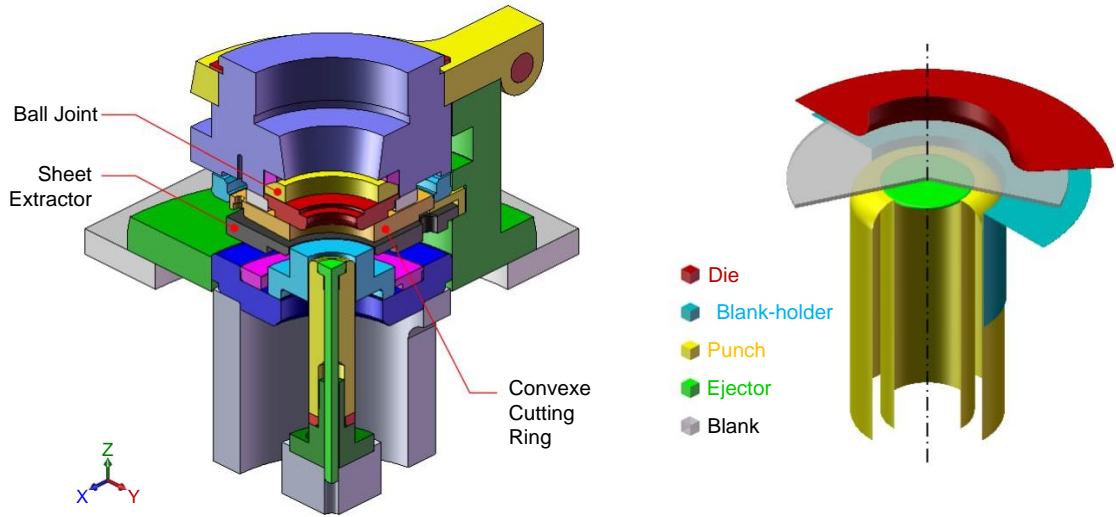

Figure 1. Schematic representation of the Swift test device (left) used on the BUP200 machine and simplified representation of the device (right).

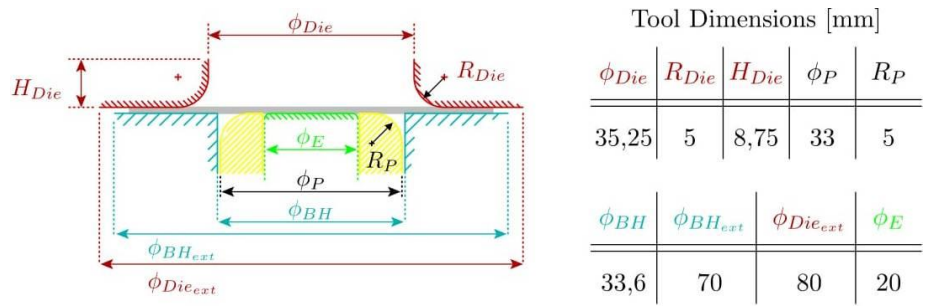

Figure 2. Schematic representation of the Swift tool including the main dimensions.

## 2.2 Experimental results

Figure 3 presents the punch force evolution with its displacement for tests performed under dry and lubricated conditions. Interrupted tests were also performed for lubricated conditions, to analyse the part geometry at every 5 mm of punch displacement. For lubricated conditions, the maximum punch force value of 17.5 kN is attained for a punch displacement of 11-12 mm, which corresponds to the instant that the die and the punch shoulder radii are completely formed in the part. As shown in Figure 2, there is no way in the tool to control the blank-holder displacement. Thus, it only stops its movement when it establishes contact with the die. This means that when the blank-holder loses contact with the blank, a sudden change is observed in the punch force evolution, for a punch displacement of approximately 19 mm. In the dry conditions test, there is a small sudden decrease of the force. For the lubricated tests, it results in a small increase of the punch force, since the blank-holder was promoting the blank flow into the die cavity. The ironing stage starts to occur for a punch displacement of approximately 21 mm. For a punch displacement of 25 mm to 30 mm small oscillations of the punch force occur, as a result of the ironing of the ears. A perfectly symmetrical test will lead to only one force oscillation, resulting from the simultaneous ironing of the four ears. Slightly asymmetrical tests present several peaks corresponding to slightly different ears. Nevertheless, the tests are very reproducible, particularly until the starting of the ironing stage, as shown in Figure 3. The dry test was performed with the tools and the blank degreased, leading to a force peak for the punch displacement of approximately 16 mm. This results from adherence between the tools and the blank, i.e. galling occurs [40]. The occurrence of this phenomenon for aluminium alloys has been previously reported as a consequence of the physicochemical interaction between the blank and the tools, for low lubricant conditions, which lead to the local heating and the adhesion of the aluminium alloy to the tools. This phenomenon tends to occur in zones subjected to high contact pressures, as for example in the die shoulder [41–43].

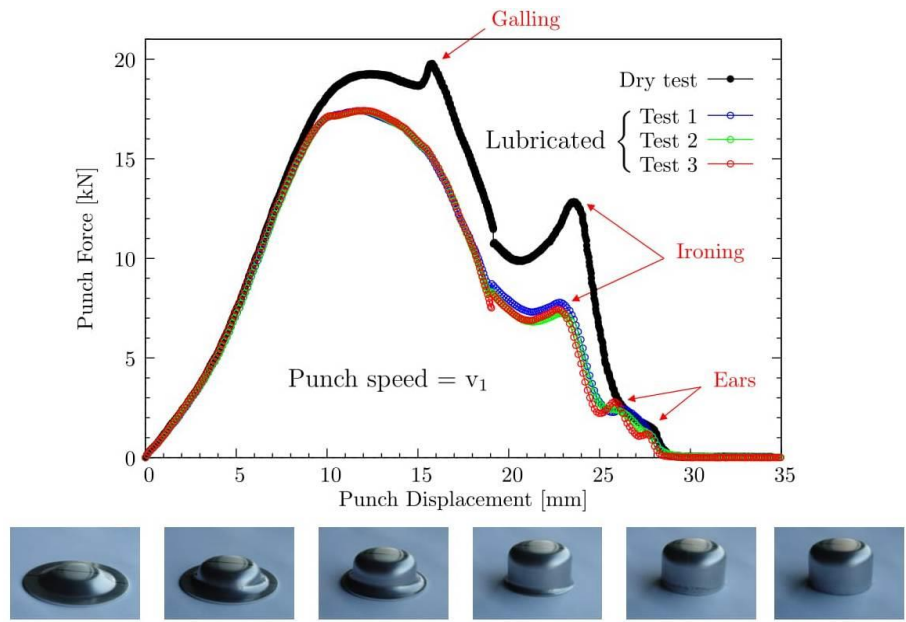

Figure 3. Punch force *versus* punch displacement for a punch velocity $v_1$, considering different lubrication conditions. Cup geometry at every 5 mm of punch displacement, obtained from the interrupted tests.

The maximum thickness value measured in the cup's flange, for a punch displacement of 20 mm (i.e. before the ironing stage), was 1.3 mm. Therefore, the theoretical thickness reduction induced by the ironing stage is 15% [29], which is a moderate value when compared with the typical values used in can making industries. The thickness measurements were performed with a (MMT) Brown&Sharpe® MicroXcel pfx 4.5.4 tri-dimensional measurement machine. The measurements were performed from the cup's bottom along three orientations: rolling direction (RD), 45º to the RD (DD) and the transverse direction to RD (TD). Figure 4 presents the results for the dry and the lubricated tests, showing an evolution similar to the ones previously reported for the Swift test, with a higher thickness reduction close to the exit of the punch shoulder radius [3], with a value that depends on the punch and die shoulder radius [44]. Nevertheless, the thickness evolution shows a small anisotropic behaviour, with the DD presenting the highest thinning value, which is coherent with the highest $r$-value and the lowest ultimate tensile strength, since the yield stress and the $r$-value have an opposite effect in the cup height [32]. The lack of lubrication only generates a more accentuated thinning of the vertical wall, which indicates that the contact conditions between the blank and the die have a strong influence on the results [45].

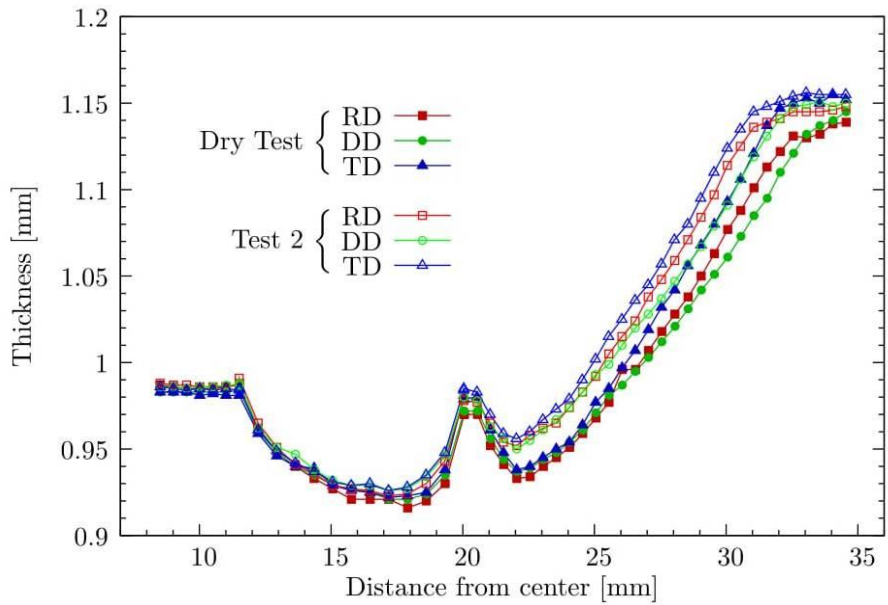

Figure 4. Thickness evolution along the curvilinear coordinate for RD, DD and TD, for a punch velocity $v_1$, considering different lubrication conditions.

The earing profile was also measured with the MMT machine and the results are presented in Figure 5, showing the good reproducibility of the measurement for the four tests performed under lubricated conditions. For dry conditions, the draw-in is smaller resulting in a slightly higher thickness reduction of the vertical wall (see Figure 4), leading to a higher cup. Whatever the lubrication conditions, it is possible to observe four ears, at 45º to RD, which corresponds to the typical behaviour of a material presenting a planar anisotropy coefficient $\Delta r = -0.170 < 0$ [3, 46]. Also, since the $r$-value along RD is smaller than along TD, the valleys at 0º with RD are less pronounced than the ones at 90º to RD. The earing profile is mainly dictated by the in-plane $r$-values directionalities because the material presents only a small variation of the flow stresses.

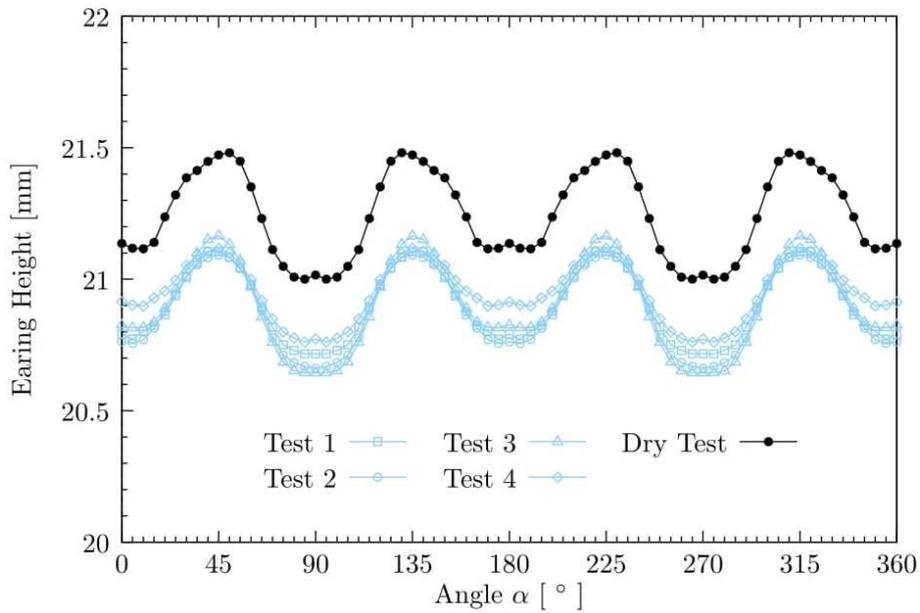

Figure 5. Earing profile after cup forming, for a punch velocity $v_1$, considering dry and lubricated conditions (four tests).

The strain fields were measured for different punch displacements using the digital image correlation system ARAMIS (ARAMIS 4M, with two video cameras and an image resolution of $2048 \times 2048$ pixels$^2$) by making an image of the initial sheet (before deformation) and an image for each instant under analysis. A thin polypropylene film (0.02 mm of thickness) was used between the die and the blank to preserve the integrity of the paint speckle, allowing the strain field calculation. The size of the correlation windows is $13 \times 13$ pixels$^2$ and the scale of the order of 18 pixels/mm. The measurement is carried out with a step size of 8 pixels corresponding to a recovery of 38%.

The interrupted tests were performed under lubricated conditions and using the polypropylene sheet between the blank and the die. The strain fields obtained allow to represent the strain states of the exterior surface of the cup in the forming limit diagram. The results are shown in Figure 6 for a punch displacement of 10 mm, 20 mm and 30 mm, which also presents the forming limit curve extracted from [4]. The points located at the cup's bottom follows a monotonous plane strain path. The points located on the vertical wall also present a more or less monotonous plane strain path ($\varepsilon_2 = 0$), since the sheet extends only in the axial direction. The strain path for the points located in the flange is approximately uniaxial compression $\varepsilon_2 = (-2\varepsilon_1)$, i.e. the sheet thickens due to the circumferential compression. Thus, a variation of the strain path occurs when the material moves along the die shoulder radius, ranging from uniaxial compression to plane strain, which is also accentuated by the ironing stage. This analysis of the strain paths and its changes is in agreement with numerical results previously reported in [28], as well as for interrupted experimental results [4], both performed for another cylindrical cup geometry.

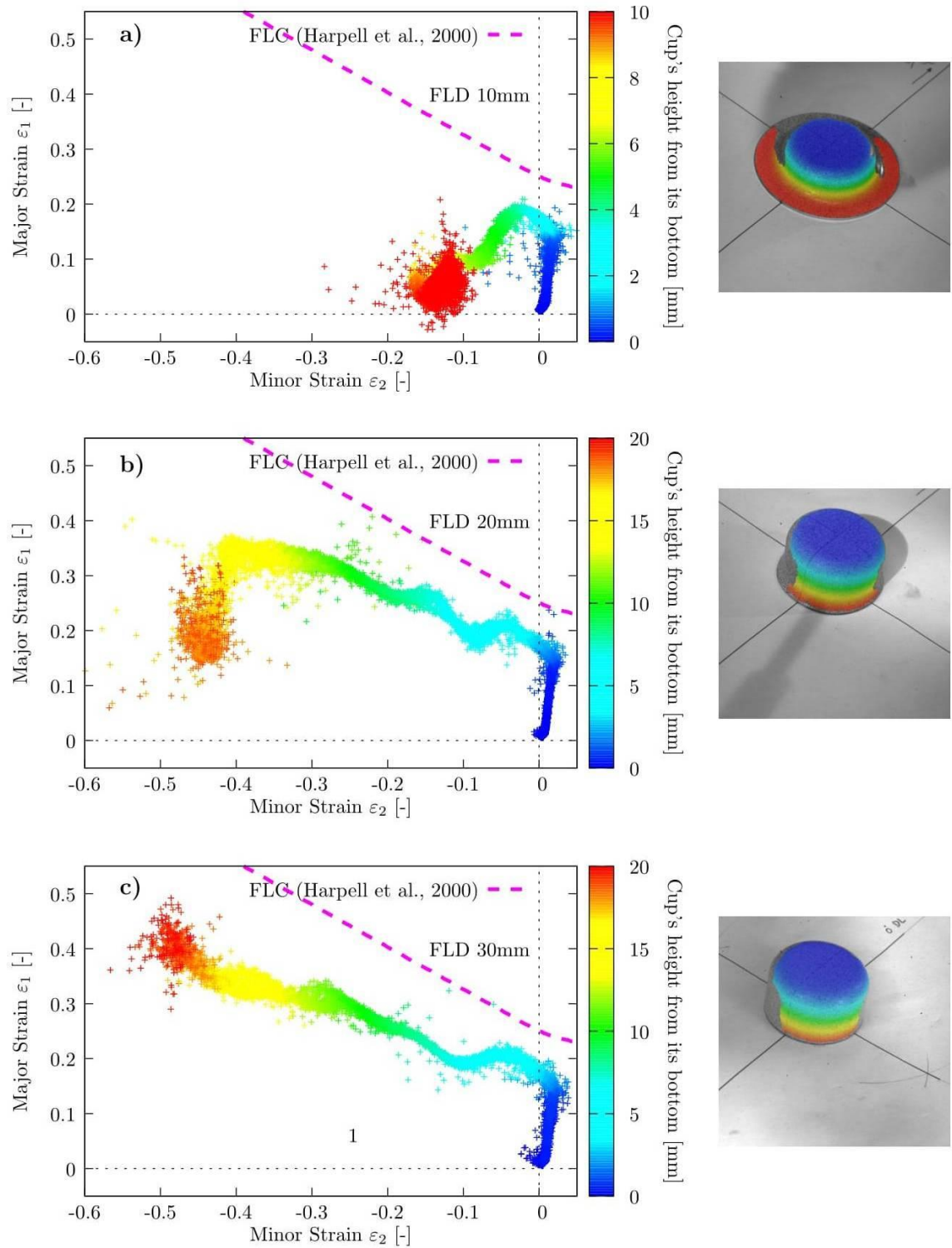

Figure 6. Major *versus* minor strain (obtained from DIC by Aramis system) for a punch displacement of: (a) 10 mm; (b) 20 mm and (c) 30 mm. The scale corresponds to the cup's height, measured from its bottom, in order to allow the correlation between the strain state and the location in the cup.

The Demeri test is used to characterize the springback as previously done by other authors for other aluminium alloys [16, 17, 47–49]. This test consists on trimming a ring in the cup's vertical wall that is afterwards cut along the RD. The ring opening allows the direct measurement of springback resulting from the release of the internal stresses [48]. The rings were trimmed at a distance of 8 mm from the cup's bottom with a height of 7 mm, using an electro-erosion machine by wire. The wire and the electric arc generate a cutting thickness of 0.3 mm, which is taken into account to obtain a ring of 7 mm in height. The same technique was used to cut the rings along the RD, as shown in Figure 7, and the opening was measured. Table 1 shows the results for three tests indicating an average value of 6.0 mm.

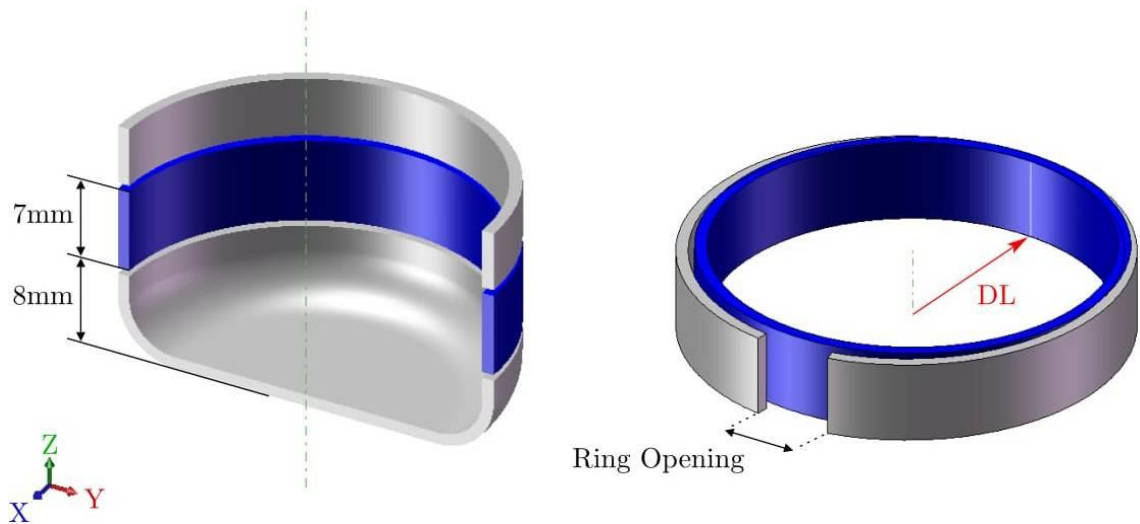

Figure 7. Positioning of the ring in the cup and measurement of its opening after springback.

Table 1. Measured values for the opening of rings cut from three cups obtained considering lubricated conditions.

| Test number  | 1         | 2         | 3         | Average   |
|--------------|-----------|-----------|-----------|-----------|
| Opening [mm] | 6.05±0.02 | 5.95±0.02 | 6.00±0.02 | 6.00±0.07 |

## 3  Numerical simulation of the Swift test

The numerical simulations were performed with the standard-implicit version of the ABAQUS. The tools were modelled using analytical surfaces since they were assumed as rigid. The die opening radius corresponds to the measured value of 35.30 mm (see Figure 2). A ring of elements called "Block stopper" was defined to establish contact with the blank-holder and prevent its movement, as soon as it loses contact with the blank sheet. This allows to reproduce the contact conditions previously mentioned. Moreover, a quarter of the model was sufficient to perform the simulation of the deep drawing process, due to the geometrical and material symmetry conditions. However, in order to simplify the simulation of the ring opening, half-model was considered.

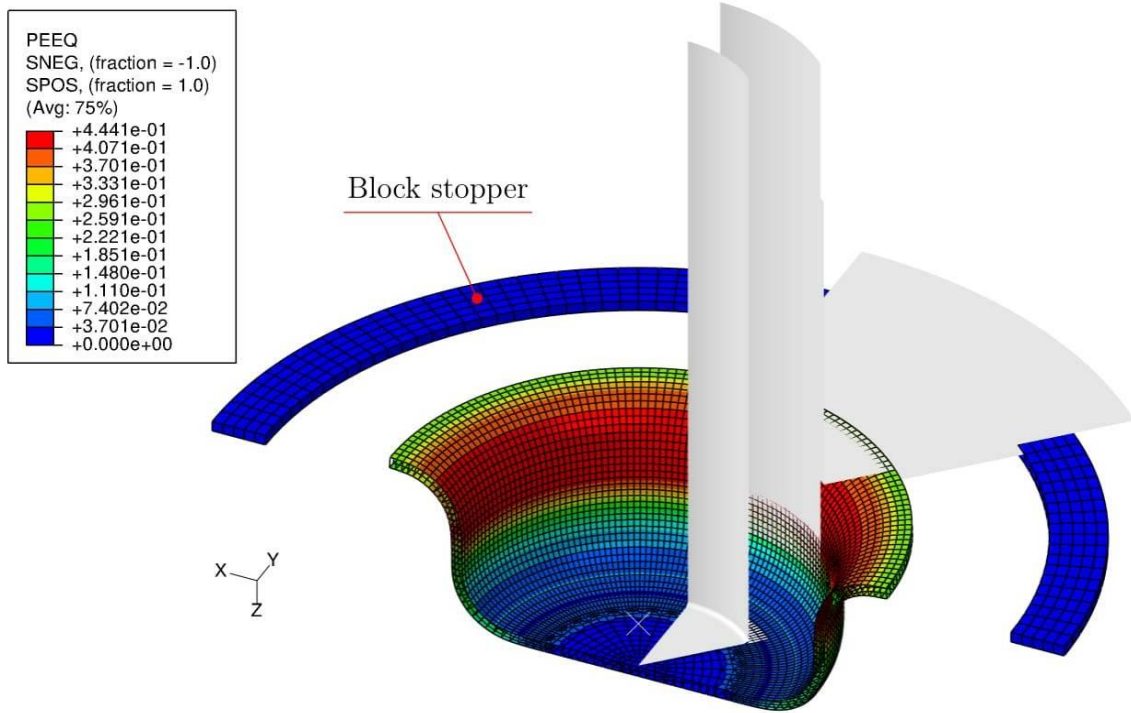

Figure 8. Model used in the numerical simulation of the cup, presenting the equivalent plastic strain distribution for a punch displacement of 15 mm.

## 3.1  Discretization of the blank

Two types of finite elements available in ABAQUS were selected to perform this study, based on previous results that show that the C3D8I is the solid element which leads to better springback predictions [16]. The C3D8I element is a linear hexahedron element (with 8 nodes) and selectively reduced integration to which incompatible deformation modes are added. The SC8R element (continuum shell) is a solid-shell, i.e. it is also a hexahedron with 8 nodes (with only translation degrees of freedom), but of shell type with respect to the kinematic behaviour. It is an element with reduced integration (only 1 integration point in the plane) for which it is possible to vary the number of integration points in the thickness direction (for further details see [50]). The Simpson integration rule was selected considering 5 (default) and 15 through-thickness points, to analyse the effect of this parameter on the springback, since it is known to have a strong impact in its prediction [14, 18, 51]. For the simulations carried out with the C3D8I element, three layers of elements were used, to have 6 integration points through the thickness.

Two meshes were used to study the influence of the in-plane mesh size, considering a different number of elements in the radial and circumferential direction, as shown in Figure 9, for the coarse mesh. The part of the mesh corresponding to the bottom of the cup is subjected to small deformation values (see Figure 6 and Figure 8) and, consequently, was considered identical for the two meshes (total of 88 elements in the plane). The coarse mesh (M1) is built considering 34 and 48 elements along the radial and circumferential directions, while the fine mesh (M2) has 68 and 96, respectively. Thus, when using SC8R element, the mesh M1 has 3984 elements (8164 nodes) and M2 has 15312 elements (30994 nodes). The meshes for the C3D8I element type has twice has many nodes.

The transition zone allows the refinement of the part which corresponds, at the end of the forming stage, to the vertical wall of the cup where the ring will be cut. The average mesh size in this part of the mesh is 0.5 and 0.25 mm for the meshes M1 and M2, respectively. Both discretizations respect the recommendations of using an element size that covers at least 5 to 10º of the tool radius, i.e. at least 9 elements in contact with the radius of the die or punch, in order to accurately predict the springback [14, 51]. On the other hand, shell elements are adequate to predict springback when the ratio between the tool radius and the thickness of the sheet is greater than 5-6, while solid elements are required for smaller ratios [14]. Previous results also indicate that, in case of solid elements, the ratio between the finite element length in the sheet plane and through-thickness should be as close to 1.0 as possible, in order to improve springback predictions [52], which justifies the selection of a small in-plane finite element size. Finally, in order to perform the simulation of the ring opening, a predefined region is established on the original mesh [16, 33, 49], as shown in Figure 9. The simulation of this test is done by deactivating the zones of the mesh not corresponding to the ring. This means that for different blank discretizations the ring can have a different height, which is known to affect the ring opening [48]. This effect is minimized with the selection of a small in-plane mesh size.

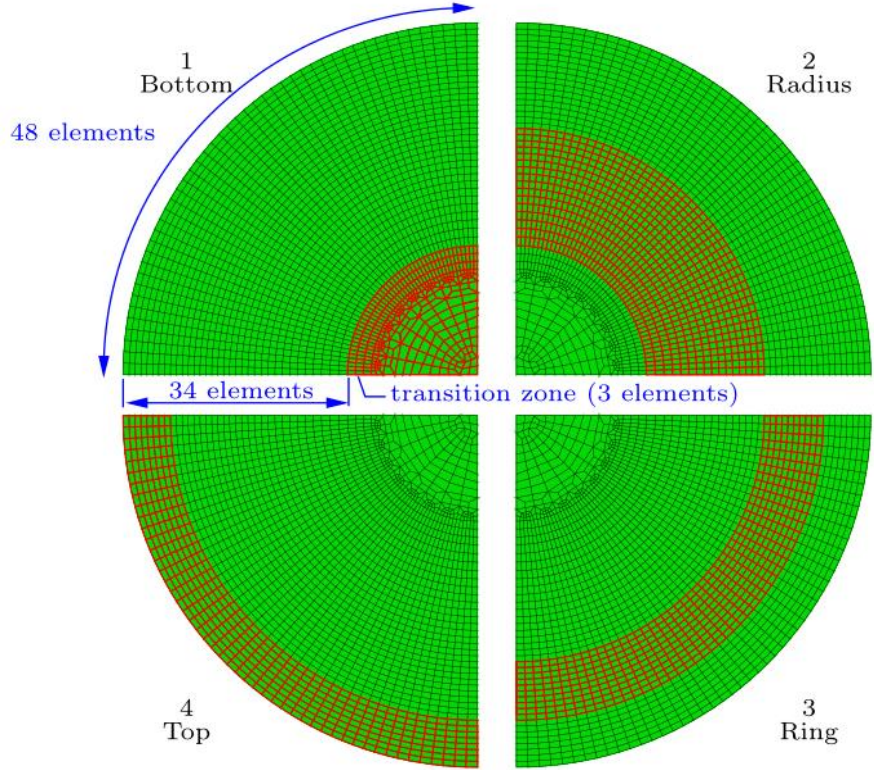

Figure 9. Predefined zones on the mesh M1 (coarse mesh) for the geometrical cutting of the ring, including the details about the definition of the mesh characteristics [33].

## 3.2 Elastoplastic constitutive model

The elastic behaviour is assumed isotropic and is described by a Poisson coefficient of 0.33 and a Young's modulus of 68 GPa. The reversed shear results showed a very weak Bauschinger effect [53] and, consequently, an isotropic hardening law was adopted. Based on the results available, it was decided to identify the hardening behaviour using the tensile test performed with the specimen aligned along the RD. However, as shown in Figure 8, the part will attain levels of equivalent plastic strain that will require the extrapolation of the mechanical behaviour predicted from the simple tensile test. Thus, two isotropic hardening laws were adopted, the Voce law [54, 55] and the Hockett-Sherby [56]

$$Y = Y_0 + Q\left\{1 - \exp(-C_y(\bar{\varepsilon}^{\mathrm{p}})^n)\right\}, \qquad (1)$$

where $\bar{\varepsilon}^{\mathrm{p}}$ denotes the equivalent plastic strain, $Y_0$ is the initial value of the yield stress, $Q = (Y_{\mathrm{sat}} - Y_0)$ where $Y_{\mathrm{sat}}$ is the flow stress saturation value and $C_y$ defines the growth rate of the yield surface. For the Voce law $n = 1$, while for the Hockett-Sherby law this parameter allows to increase the value of the flow stress saturation value. The identification of these parameters, for both laws, was performed using the least squares method, in order to minimize the difference with the results obtained for the tensile test with the specimen oriented along the RD, and the parameters obtained are presented in Table 2. Figure 10 presents the comparison between the results obtained with the Voce and the Hockett-Sherby laws and the experimental one. The difference between the two hardening laws is more clear only at the end of the experimental test and, consequently, for the extrapolated range, which means the difference will only be noticeable in the areas the cup attains equivalent plastic strain values higher than 0.2. The data generated with both laws was used to define the stress-strain curve in ABAQUS.

Table 2. Parameters identified for the Voce and Hockett-Sherby laws, for the AA5754-O aluminium alloy.

|  | $Y_0$ [MPa] | $Y_{\text{sat}}$ [MPa] | $C_y$ | $n$ |
|---|---|---|---|---|
| Voce law | 102.75 | 292.14 | 13.50 | 1.0 (imposed) |
| Hockett-Sherby law | 91.74 | 308.63 | 7.98 | 0.831 |

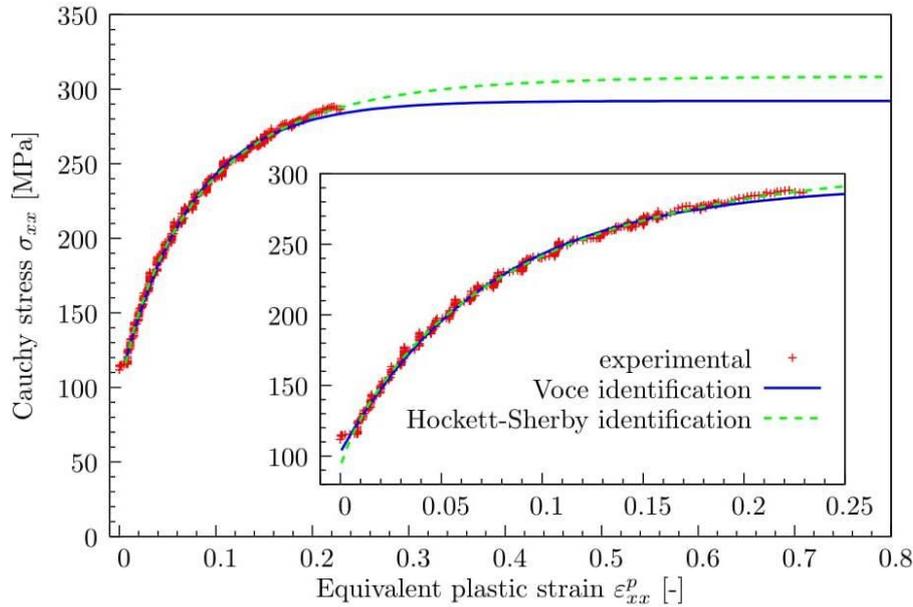

Figure 10. Comparison between the Cauchy stress - equivalent plastic deformation curves obtained using the isotropic hardening laws of Voce and Hockett-Sherby and the experimental result from the tensile test with the specimen oriented along the RD.

Since the aim of this study is to compare the computational efficiency of both type of elements, to assure the same implementation conditions, it was decided to adopt only yield criteria available as standard in ABAQUS. Thus, the orthotropic behaviour is described by the von Mises and the Hill48 [57] yield criteria (both available as standard in ABAQUS code). According to the Hill'48 yield criterion, defined in the appropriate orthogonal rotating orthotropic frame, the equivalent stress $\bar{\sigma}$ is expressed by:

$$\bar{\sigma}^2 = F(\sigma_{22} - \sigma_{33})^2 + G(\sigma_{33} - \sigma_{11})^2 + H(\sigma_{11} - \sigma_{22})^2 + \\ +2L(\sigma_{23})^2 + 2M(\sigma_{13})^2 + 2N(\sigma_{12})^2, \quad (2)$$

where F, G, H, L, M and N are the parameters that describe the anisotropic behaviour of the material, while $\sigma_{11}$, $\sigma_{12}$, $\sigma_{33}$, $\sigma_{23}$, $\sigma_{13}$ and $\sigma_{12}$ are the components of the Cauchy stress tensor defined in the orthotropic frame. The von Mises corresponds to the particular case the material presents isotropic behaviour, i.e. F = G = H = 0.5 and L = M = N = 1.5. In order to identify the orthotropic behaviour of the material, it was decided to use only the *r*-values (see Section 2.1), since it is known that this yield criterion is unable to describe both the yield stress and the *r*-values in-plane directionalities simultaneously. Moreover, since the hardening behaviour was identified based only on the tensile test performed at RD, the condition G + H = 1 was also adopted. The sheet is assumed isotropic through the thickness, leading to L = M = 1.5. In ABAQUS, the anisotropic behaviour is described based on the stress ratios $R_{ij}$, which can be obtained from the anisotropy coefficients or from the anisotropy parameters:

$$R_{11} = \frac{1}{\sqrt{G+H}} = 1; R_{22} = \frac{1}{\sqrt{F+H}} = \left(\frac{r_{90}(1+r_0)}{r_0(1+r_{90})}\right)^{0.5}; R_{33} = \frac{1}{\sqrt{F+G}} = \left(\frac{r_{90}(1+r_0)}{r_0+r_{90}}\right)^{0.5};$$

$$R_{12} = \frac{1}{\sqrt{2N/3}} = \left(\frac{3r_{90}(1+r_0)}{(2r_{45}+1)(r_0+r_{90})}\right)^{0.5}; R_{13} = \frac{1}{\sqrt{2M/3}} = 1; R_{23} = \frac{1}{\sqrt{2L/3}} = 1. \quad (3)$$

Table 3 presents the values obtained for this stress ratios, which result in an in-plane distribution of the yield stress that presents the lowest value at DD, but the highest value at TD.

Table 3. Values of the Hill48 stress ratios $R_{ij}$ identified from the anisotropy coefficients, for the AA5754-O aluminium alloy.

| $R_{11} = R_{13} = R_{23}$ | $R_{22}$ | $R_{33}$ | $R_{12}$ |
|---|---|---|---|
| 1.0 (imposed) | 1.03 | 0.93 | 0.98 |

In deep drawing operations, it is difficult to determine the friction coefficient, both experimentally and numerically. Several authors agree that its value is not constant and it will depend on the type of lubrication, the shape of the contact surfaces of the tools (flat surface or rounded matrix), the forming velocity or the contact pressure [58, 59]. From a numerical point of view, this parameter is often considered constant and determined based on trials, in order to describe as close as possible the maximum force value and the draw-in observed experimentally. This was the approach adopted in this work, although it is known that there is a strong correlation between the friction coefficient and the yield criterion adopted [60, 61].

# 4 Results and discussion

This section presents the comparison between experimental and numerical results, obtained with the model described in the previous section. Due to the wide set of numerical and material parameters under analysis, this section is organized as follows. First, the influence of the element type on the results is discussed, considering that the constitutive model is the von Mises yield criterion and the Voce hardening law. This model was the one selected also to fit the friction coefficient. Then, the influence of the constitutive model is analysed, based on the conclusions extracted from the Section 4.1. The last section analysis the influence of the die opening radius on the results, since this process parameter can have a strong impact in the ironing stage.

## 4.1 Influence of the element type

Figure 11 compares the experimental evolution of the punch force with its displacement with the results obtained using the two discretizations previously described (M1 and M2), for both the solid-shell element (Figure 11 (a)) and the solid element (Figure 11 (b)). The results are shown for the two values of friction coefficient which allow a better description of the maximum force for the drawing stage ($\mu = 0.06$) and the ironing of the wall stage ($\mu = 0.09$). The results show that the type of element adopted mainly affects the evolution of the force in the ironing stage. The in-plane refinement of the mesh seems to have a smoothing effect on the results for both the drawing and the ironing stages. The oscillations observed in the ironing stage, which are more pronounce for the coarse mesh (M1), correspond to the successive movement of the elements into the vertical part of the die. Globally, the C3D8I element presents higher force values than the SC8R element at the ironing stage, which can be related with higher thickening values predicted for the flange during the drawing stage.

Figure 12 presents the thickness evolution along the cup's wall, with each type of element and mesh M1, as a function of the friction coefficient ($\mu$). The comparison with the experimental results indicates a good correspondence with the numerical ones, although the SC8R element predicts the thinning zones less accurately, which can be related with the use of a single layer of elements in the thickness direction. Moreover, although not shown here, for the SC8R element, the number of integration points through the thickness has almost no influence on the results for the punch force and the thickness evolution.

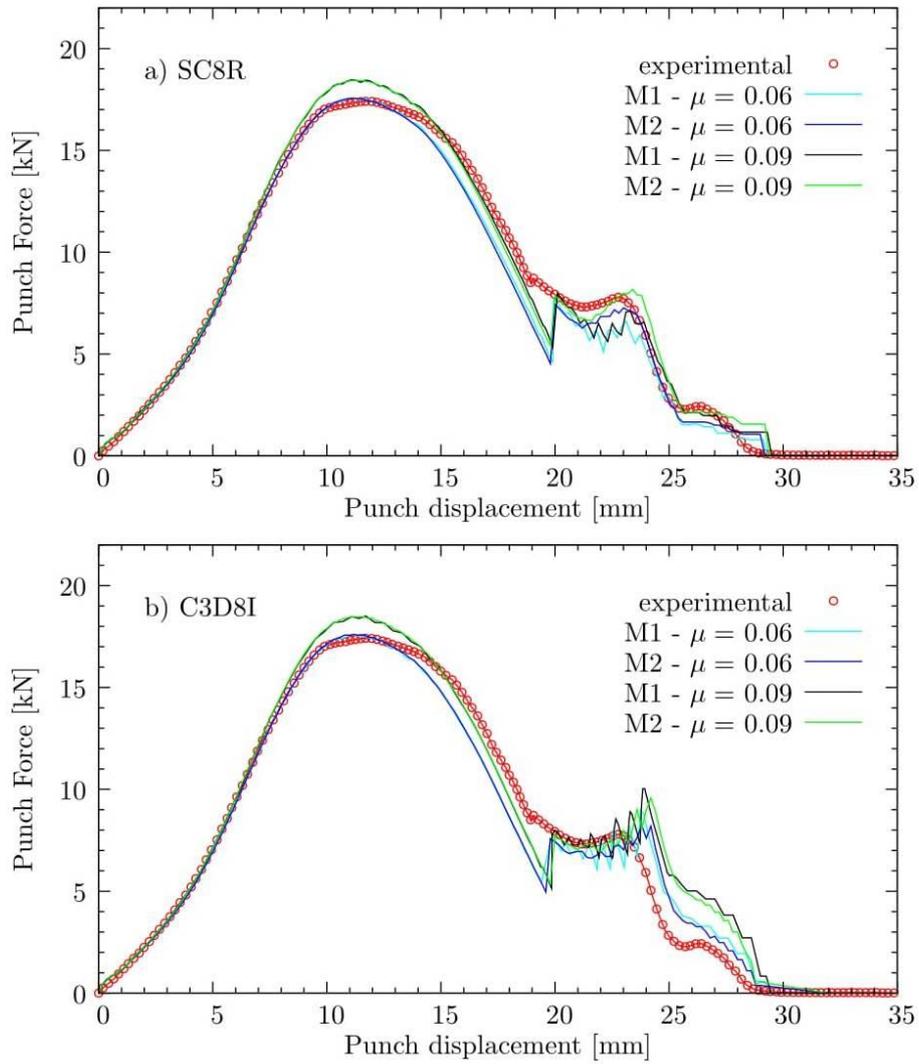

Figure 11. Evolution of the punch force with its displacement as a function of the mesh and the type of finite element adopted (results obtained with the von Mises yield criterion and the Voce hardening law).

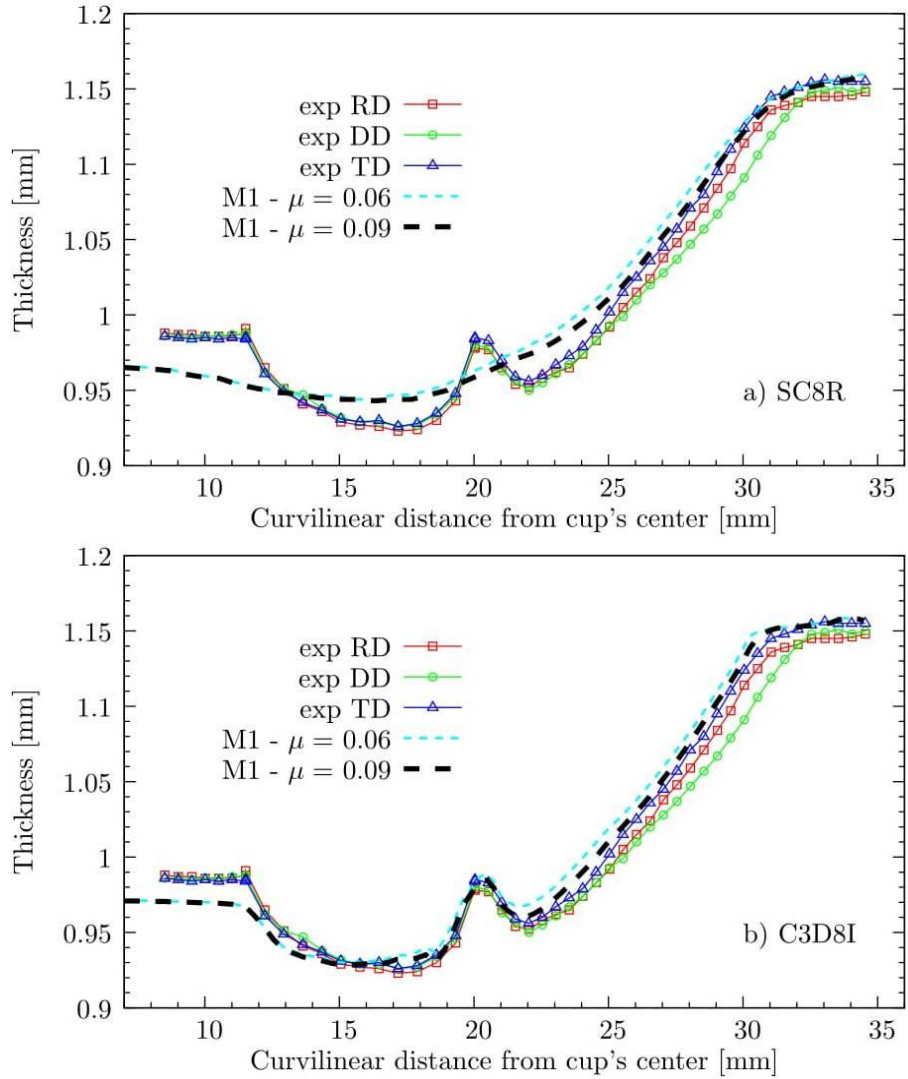

Figure 12. Thickness evolution along the curvilinear coordinate as a function of type of finite element adopted (results obtained with mesh M1 and the von Mises yield criterion and the Voce hardening law).

Figure 13 presents the ring opening predicted using the different types of finite elements, highlighting the impact of the number of integration through the thickness in case of the SC8R element. In fact, for this type of element the increase of the number of integration points through the thickness tends to decrease the springback value predicted for a given mesh, indicating a more rigid behaviour for a small number of integration points. On the other hand, the finer in-plane discretization leads to higher springback values, whatever the type of finite element adopted. The same applies to the friction coefficient, i.e. higher values lead to a larger ring opening.

It should be mentioned that the increase in the cup's average height with the increase of the friction coefficient can also contribute to the increase of the ring opening.

In fact, as shown in Figure 9, the numerical model adopted always assumes the same set of finite elements to define the ring. Thus, the increase of the friction coefficient also contributes to the increase of the ring height. This seems to be the main effect contributing to the increase of the springback prediction, when using a higher friction coefficient value. The ring height has been previously reported as one of the factors contributing for changes in the springback value [48].

The numerical ring opening predictions are all close to the experimental mean value of 6 mm. However, the different in-plane discretizations and the two types of elements tested generate quite different computational times. A factor of about 4.6 is obtained between the simulation times of the meshes M1 and M2. For the two types of finite elements tested and for a given mesh, this factor is 4.15 between the SC8R and C3D8I elements. Therefore, the SC8R element is more cost-effective, since all the experimental results are correctly predicted, with a much smaller computational time. However, the thickness prediction is more accurate when using the C3D8I element.

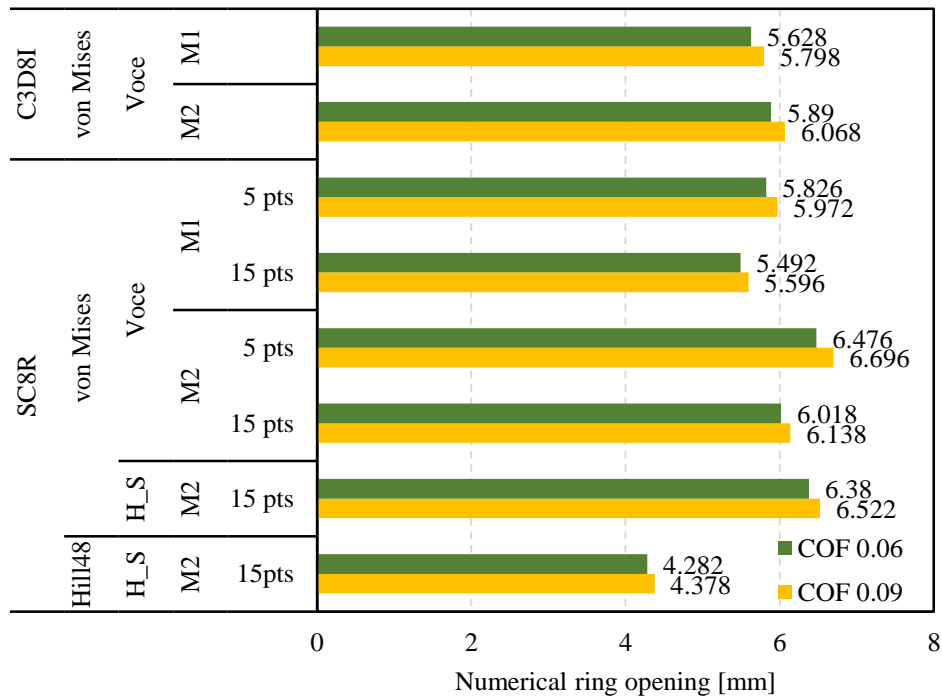

Figure 13. Ring opening predicted as a function of type of finite element, the in-plane discretization, the yield criteria, the friction coefficient (COF) and the hardening law. The label "H_S" corresponds to the Hockett-Sherby hardening law.

## 4.2 Influence of the constitutive model

The analysis of the influence of the yield criteria adopted on the numerical results of an AA5754-O cylindrical cup drawing, including the ring opening, has been previously performed for the C3D8I finite element [16], but for another cup dimensions. In this study, the analysis was performed considering the von Mises, the Hill48 and the non-quadratic yield criterion, Barlat91, proposed in [62]. Globally, the results show that the Hill48 yield criterion overestimates the thickness distribution, while the Barlat91 accurately describes it. However, regarding the springback results, the von Mises yield criterion predicts the ring opening value closer to the experimental one. In fact, both orthotropic yield criteria lead to a clear underestimation of the ring opening, particularly the Hill48 [16]. Although the cup dimensions analysed in this previous study are different from the ones presented here, the same conclusions are valid for the C3D8I finite element.

Based on these previous results, the analysis of the influence of the hardening law and the yield criterion will be performed considering the M2 mesh and the SC8R element, with 15 integration points through-thickness. Figure 14 compares the experimental evolution of the punch force with its displacement with the results obtained considering the Hockett-Sherby hardening law and the two yield criteria under analysis. Regarding the hardening law, the comparison of the results of Figure 14 with the ones shown in Figure 11 (a) indicates an excellent correlation with the experimental results up to 12 mm of punch displacement, for the two hardening laws, with a coefficient of friction of 0.06. In fact, the differences in the punch force evolution predicted with both hardening laws are negligible. Nevertheless, with the Hockett-Sherby law, a better prediction of the maximum force for the ironing stage is attained with a friction coefficient of 0.06. On the other hand, regarding the numerical prediction of the ring opening, shown in Figure 13, the Hockett-Sherby hardening law overestimates the experimental value. These results are consistent with the ones in Figure 10, which highlights that for equivalent plastic strains higher than 0.2 the Hockett-Sherby law predicts higher flow stress values, which can lead to an increase of the circumferential stresses predicted for the cup wall, increasing the springback prediction.

Regarding the yield criterion adopted, Figure 14 shows that taking into account the anisotropic behaviour of the sheet mainly modifies the prediction for the ironing stage forces. This can be explained by a change in the thickness evolution along the cup's wall, since the use of the Hill48 yield criterion leads to an overestimation of the thickening, as shown in Figure 15. Nevertheless, it is observed that the order of the numerical curves along the orientations RD, DD and TD respect the order observed experimentally. This can be related with the fact that the identified parameters for the Hill48 yield criterion enable a proper description of the *r*-values in-plane directionalities.

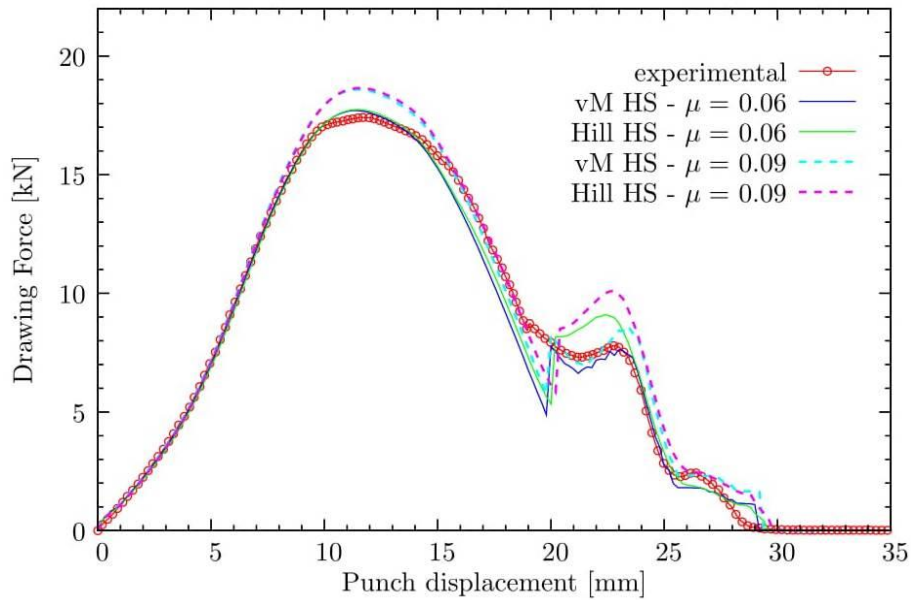

Figure 14. Evolution of the punch force with its displacement as a function of the yield criterion (results obtained with mesh M2, the SC8R-15 points and the Hockett-Sherby hardening law).

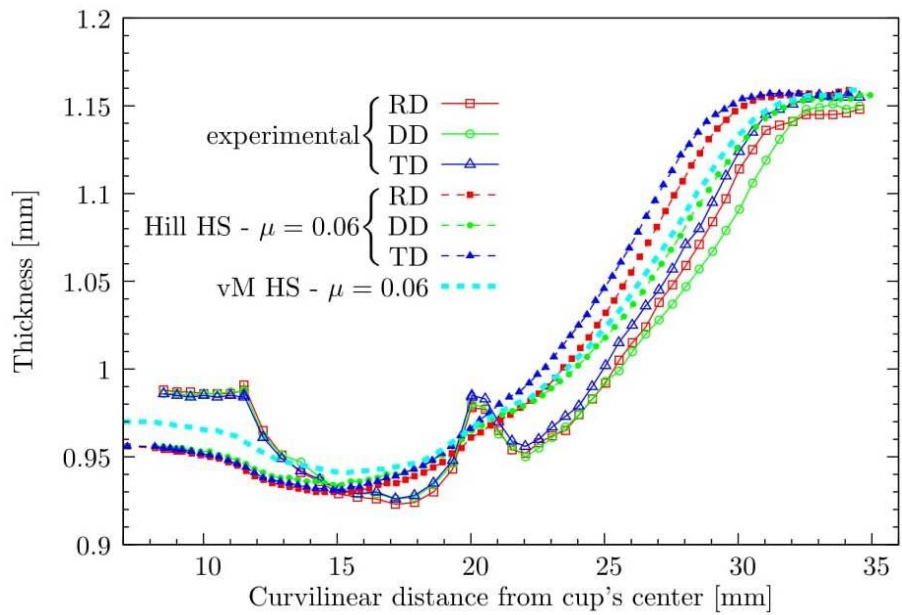

Figure 15. Thickness evolution along the curvilinear coordinate as a function of the yield criterion (results obtained with mesh M2, the SC8R-15 points and the Hockett-Sherby hardening law).

Figure 16 shows the evolution of the external diameter of the cup as a function of the punch displacement, thus reflecting the draw-in of the sheet during the forming process. The measurements of the external diameter, along the RD, were performed with a calliper using the interrupted tests (see Figure 3, labelled Classic Cups) and from the ARAMIS measurements (see Figure 6, labelled Aramis Cups). The numerical results are in agreement with the experimental ones. The Hill48 yield criterion predicts a slightly higher draw-in than the von Mises isotropic yield criterion. Also, the influence of the friction coefficient is negligible. It is also observed that the von Mises criterion better predicts the evolution of the exterior diameter over the first fifteen millimetres of punch displacement, whereas the Hill48 criterion becomes more accurate after 20 mm of punch displacement, whatever the friction coefficient adopted. Nevertheless, it should be mentioned that both yield criteria overestimate the punch displacement corresponding to the loss of contact between the blank and the blank-holder (see Figure 14), although the predicted cup average height is smaller than the experimental one.

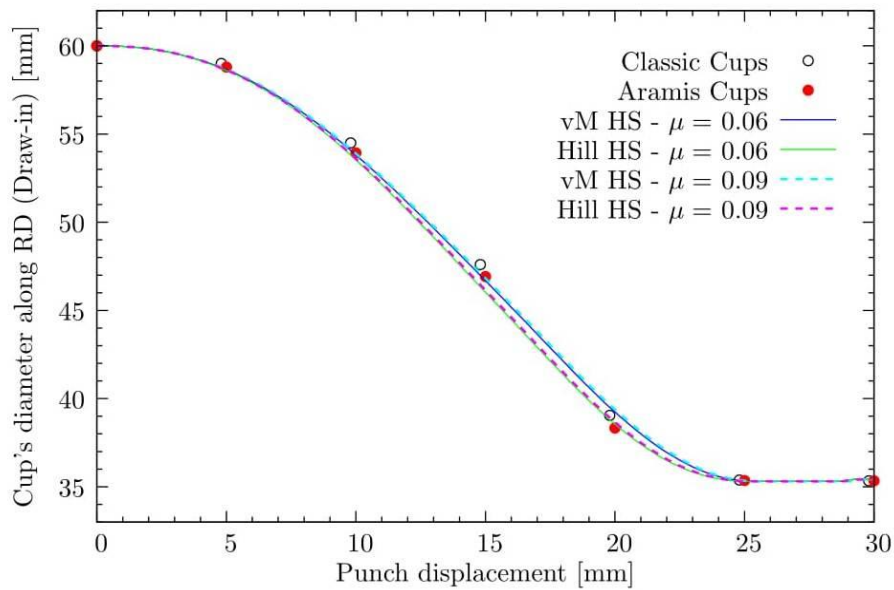

Figure 16. Evolution of the exterior diameter of the cup with the punch displacement as a function of the yield criterion (results obtained with mesh M2, the SC8R-15 points and the Hockett-Sherby hardening law).

The results obtained for the earing profile, at the end of the forming process, are shown in Figure 17, highlighting that the Hill48 yield criterion accurately predicts the position of the ears, although the amplitudes are overestimated. The overestimation of the ears amplitude can be explained by the overestimation of the in-plane yield stress directionalities, which typically occurs when the Hill48 anisotropy parameters are identified based on the *r*-values. It is known that by adjusting the Hill48 anisotropy parameters using the yield stresses, instead of the *r*-values, reduces the amplitude predicted for the ears, but the thickness prediction will be less accurate (see e.g. [28]). However, the cup average height is slightly underestimated by both yield criterion, although the increase of the friction coefficient leads to an increase of the average cup's height, as also reported in the experimental results (see Figure 5). Also, although the *r*-value predicted along RD is smaller than along TD, since the yield stress value predicted along TD is higher than along RD, the valleys at 0º with RD are similar to the ones at 90º to RD.

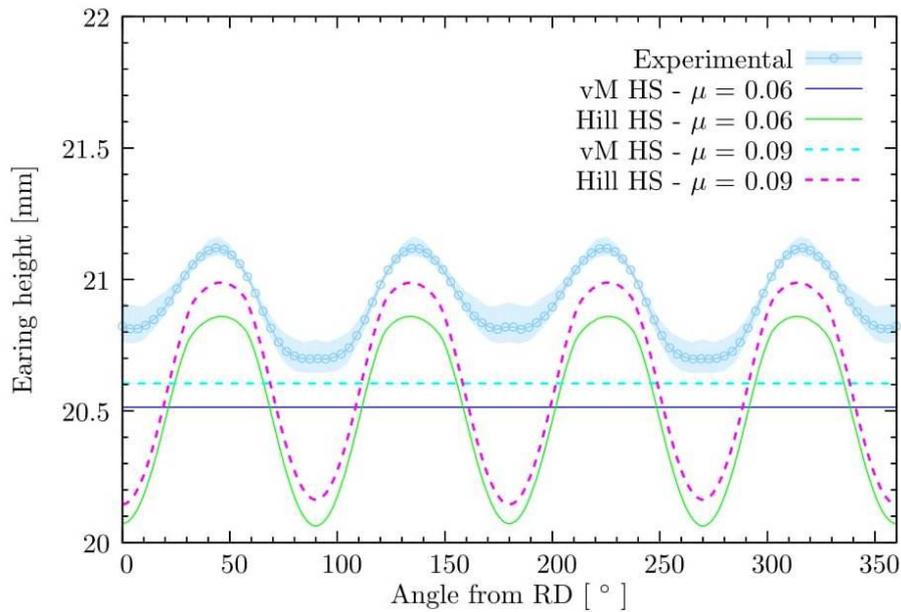

Figure 17. Earing profile as a function of the yield criterion (results obtained with mesh M2, the SC8R-15 points and the Hockett-Sherby hardening law).

Regarding the prediction of the ring opening, Figure 13 shows that the Hill48 criterion is unable to accurately describe the springback, which can be related with the inaccurate description of the through-thickness distribution of the circumferential stress in the ring. This has been attributed to the fact that the Hill48 yield criterion is known for not being able to properly describe the orthotropic behaviour of materials with average *r*-value lower than 1, particularly for equibiaixial and plane strain states [15, 16]. On the other hand, the von Mises isotropic criterion predicts results in agreement with the experimental ones, but is unable to predict the earing profile. Although not shown here, the comparison of the circumferential stress predicted in the ring for both yield criteria indicates that the introduction of the anisotropic behaviour has a strong impact in its distribution along the circumferential direction, strongly reduction the through-thickness stress gradient along the DD direction. Thus, the overestimation of the earing profile (see Figure 17) seems to directly contribute to the underestimation of the springback. However, as previously mentioned, the adoption of a non-quadratic yield criterion also leads to an underestimation of the ring opening [16], which indicates that other factors are affecting the predictions.

It is known that, after applying a prestrain value in a uniaxial tension test, the unloading modulus, measured as the slope of a secant line between the starting and end points of the unloading curve, is lower than the physically-measured Young's modulus. Several models have been proposed to take into account this phenomenon, considering its evolution with prestrain (see e.g. [63, 64]) or even describing the hysteresis loop exhibited in a load/unloading cycle (see e.g. [65]). The adoption of a model enabling the description of the degradation of the elastic stiffness due to plastic straining enables more accurate springback predictions (see e.g. [63]). Although it was not taken into account, the AA5754 alloy under analysis is known to be sensitive to this phenomenon [65]. As previously mentioned, the ring is located in a part of the cup wall which attains high levels of equivalent plastic strain. Therefore, it is reasonable to assume that the ring will present degradation of the elastic stiffness due to plastic straining. Assuming a constant reduced value for the Young's modulus, over the ring, would result in higher springback predictions, for all yield criteria.

Figure 18 (a) shows the plot of the major and minor strains obtained for the outer surface of the cup, at the end of the forming process, when using the von Mises yield criterion and the Voce hardening law. This figure highlights the strain state for each of the predefined zones (see also Figure 9), in order to improve the analysis of Figure 18 (b). This figure compares the major and minor strains obtained for the outer surface of the cup from the experimental field measurements with the results obtained with the two yield criteria. Globally, the final strain state is fairly well predicted by the two criteria, which is also an indication that the incorrect prediction of the springback by the Hill48 yield criterion should result from a poor estimation of the distribution of the internal stresses in the ring. The major difference between the numerical and the experimental results is observed in the punch radius (zone 2), where a local minimum value of the major strain is observed for a strain path close to uniaxial tension. Also, it is observed that the zone corresponding to the ring (zone 3), reaches major and minor strains higher than 0.2, which justifies the use of experimental tests that enable the description of the hardening behaviour for equivalent plastic strain values higher than the ones attained in the uniaxial tensile test.

In a previous work, considering the deep drawing of a cylindrical cup with different dimensions, for the same aluminium alloy, a similar result was observed for the strain predicted with the Hill48 and the von Mises yield criteria. Nevertheless, in that case both under-predicted the major strains in the cup wall region by about 5% strain, which was attributed to the inability of the four-node quadrilateral shell elements to accurately capture the ironing behaviour [4]. The results presented in this work confirm that the use of 3D elements provide a better prediction of the compressive through-thickness and tensile major strains in the wall. Moreover, it is interesting to note that, although the Hill48 is known for not being able to reproduce the so-called anomalous yield response for materials presenting an average $r$-value lower than 1, the strain field predicted for the outer surface is similar to the one predicted with the von Mises yield criterion. This can be related with the small earing effect observed on the cups (see Figure 17) as well as the small differences in the thickness distribution (see Figure 15), which result in a small dispersion of the trend obtain with the Hill48 yield criterion around the results predicted with the von Mises.

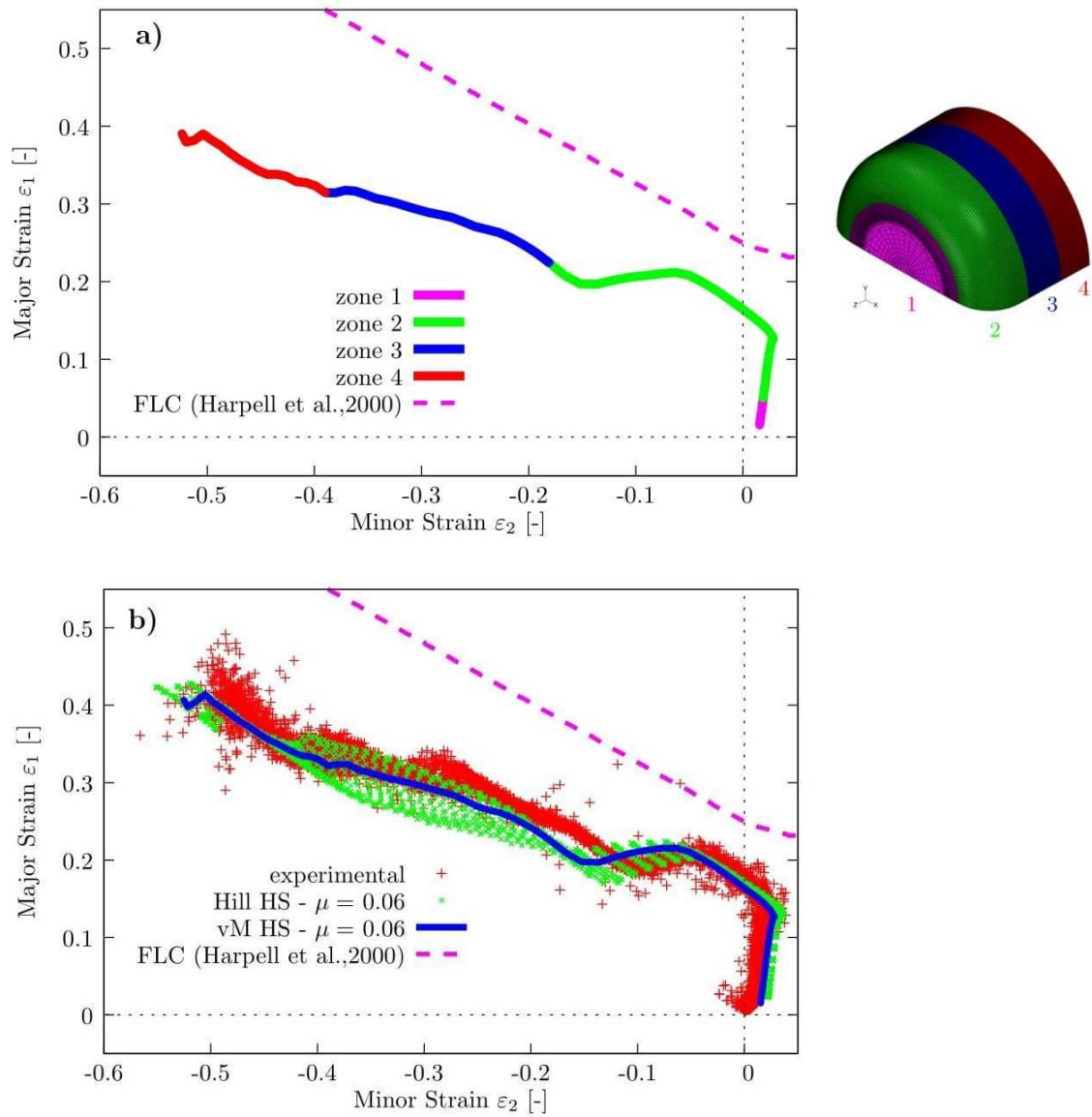

Figure 18. Major *versus* minor strain for a punch displacement of 30 mm (mesh M2 and the SC8R-15 points). Results obtained with a coefficient of friction of 0.06 and: (a) the von Mises yield criterion and the Voce hardening law; and (b) the Hockett-Sherby hardening law and the von Mises and Hill48 yield criterion.

## 4.3 Influence of the die opening radius

The geometric dimensions of the tools have a direct impact on the final shape of the manufactured parts. Thus, even small changes in the dimensions, due to tool wear or machining tolerances can influence the experimental results. In this subsection, the influence of the die opening diameter on the numerical results is analysed, using the M2 mesh, the von Mises yield criterion, the Hockett-Sherby hardening law and a friction coefficient of 0.06.

Figure 19 shows the experimental results obtained for two different die diameters. The first matrix, with diameter $\phi_1$ = 35.30 mm, corresponds to the diameter used up to now (measured on the die) and the second, diameter $\phi_2$ = 35.25 mm, corresponds to the theoretical value. This figure shows that a difference of 5 hundredths of a millimetre in the die opening diameter leads to an increase of 3 kN in the experimental maximum force attained during the ironing stage. This figure also compares the results obtained with both type of finite elements, SC8R and C3D8I. It is observed that the numerical predictions are similar up to the beginning of the ironing stage, whatever the die opening diameter used. In the ironing stage, the punch force evolution seems to be better predicted when using the SC8R element.

The results for the ring opening are presented in Figure 20, showing that the influence of the die opening diameter is negligible, which indicates that the small difference in the ironing stage of the process has little impact on the through-thickness circumferential stress distribution. However, it should be noted that the ring is cut at a height that is not submitted to ironing, i.e. according to Figure 4 only the last 5 mm are submitted to strong compression in the thickness direction, which are not part of the ring (see Figure 7). Thus, the through-thickness circumferential stress distribution is mainly dictated by the material flow on the die shoulder radius, which remained unchanged, leading to similar punch force evolutions during the drawing stage, for both die opening diameters.

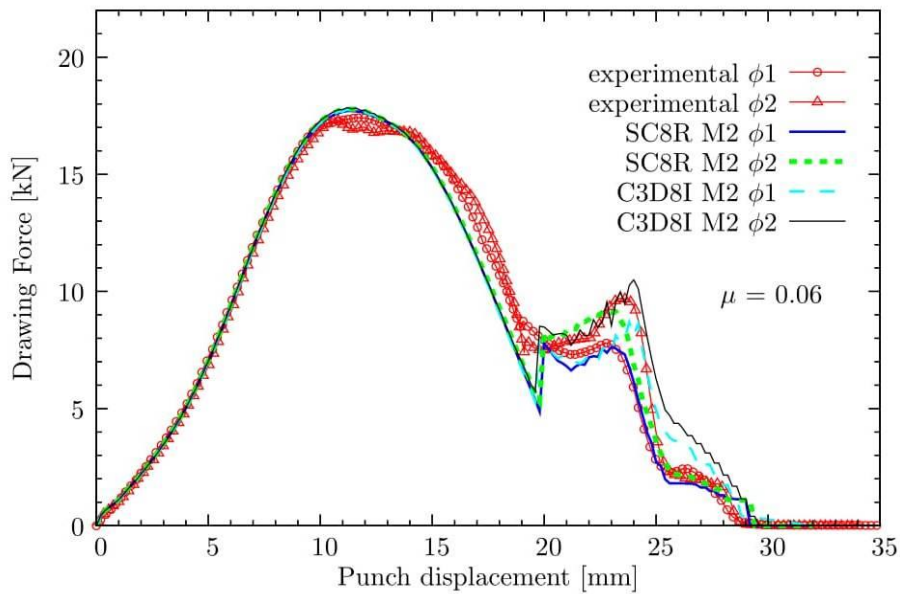

Figure 19. Evolution of the punch force with its displacement as a function of the die opening diameter (results obtained with mesh M2, the von Mises yield criterion and the Hockett-Sherby hardening law).

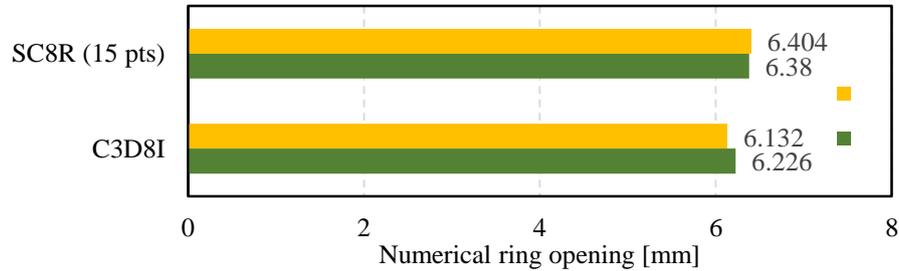

Figure 20. Ring opening predicted as a function of the die opening diameter (results obtained with mesh M2, the von Mises yield criterion, $\mu = 0.06$ and the Hockett-Sherby hardening law).

The die opening diameter has also some influence on the computational time, since a smaller value leads to some convergence difficulties associated with higher stretch values and, consequently, the computational time increases. Also, when using the Hill48 yield criterion, which generates different thickness distributions, it seems that the management of the contact is more complex, contributing to an increase of the computational time. For the SC8R element a factor of about 1.5 is obtained when the orthotropic behaviour is included in the numerical simulation, while for the C3D8I element this factor can attain a value of 3.0. As previously mentioned, the SC8R element, for which the simulations were carried out with a single element through-thickness, is much more advantageous in terms of computation time than the C3D8I element, using 3 layers of elements through-thickness.

Nevertheless, it should be mentioned that the C3D8I element allows a finer analysis of the results, as shown in Figure 12 for the thickness distribution along the cup wall. Moreover, due to the absence of a blank-holder stopper in the tool, the inner edge of the flange of the cup is pinched between the die and the blank-holder, just before the loss of contact with this tool. Due to the orthotropic behaviour of the material, this occurs mainly close to the DD, were the ears are located (see Figure 17), as shown in Figure 21. This results is well predicted by the model with C3D8I finite elements, while the one with SC8R elements is unable to predict it, due to the fact that only one through-thickness element is used, which makes it impossible to reproduce this type of distortion at the edge (linear interpolation of the displacements). This may also contribute to explain the higher computational times attained when using the C3D8I element. This pinching effect can also contribute to exaggerate any misalignment between the sample and the tools, modifying the shape of the ears [21].

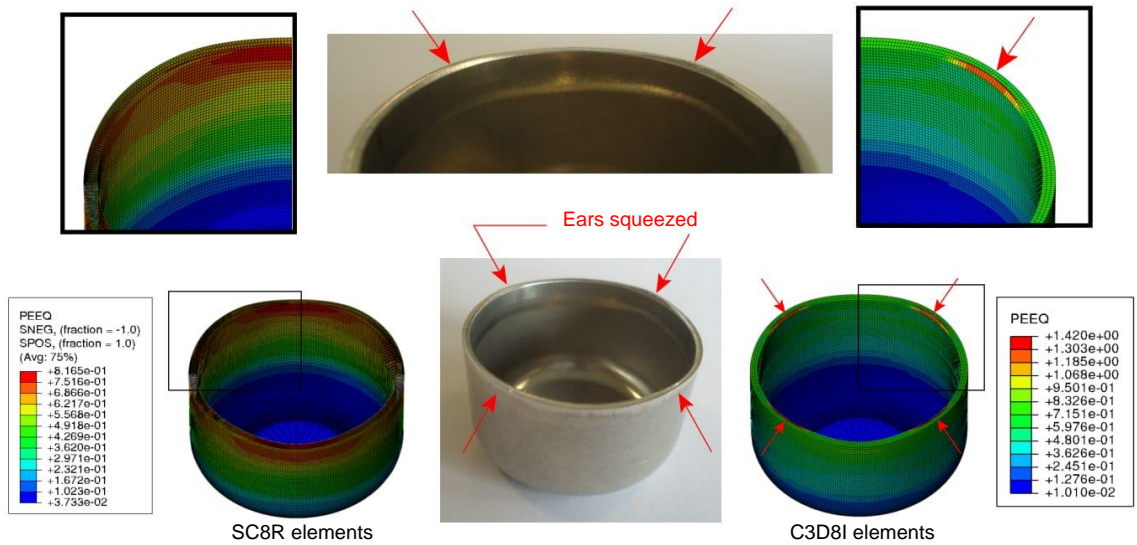

Figure 21. Detailed view of the earing profile, including the squeezing of the inner edge of the flange of the cup, close to the DD. Comparison with the numerical results obtained with the SC8R and the C3D8I elements.

## 5 Conclusions

This study presents a detailed experimental and numerical analysis of the Swift cylindrical cup forming and the subsequent split-ring Demeri test. Since the test conditions selected involve a deep drawing and an ironing stage, the use of shell elements was discarded. Two types of finite elements were selected to perform this study: the C3D8I linear hexahedron element with selectively reduced integration to which incompatible deformation modes are added; and the SC8R solid-shell with reduced integration (only 1 integration point in the plane), combined with the Simpson integration rule, considering 5 and 15 through-thickness points.

Concerning the in-plane discretization, it was observed that by guaranteeing that the discretization respects the recommendations of using an element size that covers 5 to 10º of the tool radius the punch force evolution is well predicted, although a finer discretization leads to a smoother description of the punch force in the ironing stage of the process and to an increase in the springback value predicted, whatever the type of finite element selected. The SC8R element is more cost-effective, since all the experimental results are globally accurately predicted, with a much smaller computational time (factor of 4.15 between the SC8R and C3D8I elements). However, the use of only one element through-thickness makes it impossible to accurately predict the thinning zones.

Regarding the constitutive model, the yield criterion presents a strong impact in the punch force predicted during the ironing stage, since the description of the anisotropic behaviour of the material with the Hill48 leads to higher thickening values in the flange area, when compares with the von Mises yield criterion. Although the Hill48 criterion is known for not being able to properly describe the orthotropic behaviour of materials with average *r*-value lower than 1, it correctly predicts the position of the ears as well as the trend for the thicknesses distribution along the three orientations studied. This results from the accurate description of the *r*-values in-plane directionalities, combined with a trend similar to the experimental one for the yield stress in-plane directionalities. However, since the Hill48 yield criterion overestimates the yield stress in-plane directionalities, the amplitude of the predicted ears is higher than the experimental one. This generates a gradient in the distribution of the circumferential stress along the circumferential direction for the ring that, together with an improper definition of the yield *locus* for the plane strain state, results in the underestimation of the springback, while the von Mises yield criterion leads to accurate results. Globally, both yield criteria predict the strain distributions in the final cup quite accurately. Although the strain paths and the change that occurs when the material moves along the die shoulder radius, ranging from uniaxial compression to plane strain, is well predicted, improved knowledge concerning the stress states is still required. In particular, considering that the ring opening is also underestimated when using C3D8I finite elements and a non-quadratic yield function [16], the analysis of the influence of the degradation of the elastic stiffness or the Young's modulus reduction on springback prediction, should be performed.

## Acknowledgements

The authors gratefully acknowledge the financial support of the Brittany Region (France), the Portuguese Foundation for Science and Technology (FCT), under

# References

1. Banabic D (2010) Sheet Metal Forming Processes. Springer. doi: 10.1007/978-3-540-88113-1

2. Marciniak Z, Duncan JL, Hu SJ (2002) Mechanics of Sheet Metal Forming. Butterworth-Heinemann

3. Pearce R (1991) Sheet metal forming. Springer

4. Harpell ET, Worswick MJ, Finn M, et al (2000) Numerical prediction of the limiting draw ratio for aluminum alloy sheet. J Mater Process Technol 100:131–141. doi: 10.1016/S0924-0136(99)00468-9

5. Verma RK, Chandra S (2006) An improved model for predicting limiting drawing ratio. J Mater Process Technol 172:218–224. doi: 10.1016/j.jmatprotec.2005.10.006

6. Leu D-K (1999) The limiting drawing ratio for plastic instability of the cup-drawing process. J Mater Process Technol 86:168–176. doi: 10.1016/S0924-0136(98)00307-0

7. Narayanasamy R, Ponalagusamy R, Raghuraman S (2008) The effect of strain rate sensitivity on theoretical prediction of limiting draw ratio for cylindrical cup drawing process. Mater Des 29:884–890. doi: 10.1016/j.matdes.2006.05.014

8. Zadpoor AA, Sinke J, Benedictus R (2008) Finite element modeling and failure prediction of friction stir welded blanks. Mater Des 30:1423–1434. doi: 10.1016/j.matdes.2008.08.018

9.  Marretta L, Ingarao G, Di Lorenzo R (2010) Design of sheet stamping operations to control springback and thinning: A multi-objective stochastic optimization approach. Int J Mech Sci 52:914–927. doi: 10.1016/j.ijmecsci.2010.03.008

10. Choi MK, Huh H (2014) Effect of Punch Speed on Amount of Springback in U-bending Process of Auto-body Steel Sheets. Procedia Eng 81:963–968. doi: 10.1016/j.proeng.2014.10.125

11. Alves de Sousa RJ, Correia JPM, Simões FJP, et al (2008) Unconstrained springback behavior of Al–Mg–Si sheets for different sitting times. Int J Mech Sci 50:1381–1389. doi: 10.1016/j.ijmecsci.2008.07.008

12. Lee MG, Kim SJ, Wagoner RH, et al (2009) Constitutive modeling for anisotropic/asymmetric hardening behavior of magnesium alloy sheets: Application to sheet springback. Int J Plast 25:70–104. doi: 10.1016/j.ijplas.2007.12.003

13. Schwarze M, Reese S (2011) A reduced integration solid-shell finite element based on the EAS and the ANS concept-Large deformation problems. Int J Numer Methods Eng 85:289–329. doi: 10.1002/nme.2966

14. Li KP, Carden WP, Wagoner RH (2002) Simulation of springback. Int J Mech Sci 44:103–122. doi: 10.1016/S0020-7403(01)00083-2

15. Grèze R, Manach PY, Laurent H, et al (2010) Influence of the temperature on residual stresses and springback effect in an aluminium alloy. Int J Mech Sci 52:1094–1100. doi: 10.1016/j.ijmecsci.2010.04.008

16. Laurent H, Greze R, Oliveira MC, et al (2010) Numerical study of springback using the split-ring test for an AA5754 aluminum alloy. Finite Elem Anal Des. doi: 10.1016/j.finel.2010.04.004

17. Demeri MY, Lou M, Saran MJ (2000) A Benchmark Test for Springback

Simulation in Sheet Metal Forming. doi: 10.4271/2000-01-2657

18. Wagoner RH, Li M (2007) Simulation of springback: Through-thickness integration. Int J Plast 23:345–360. doi: 10.1016/j.ijplas.2006.04.005

19. Jain M, Allin J, Bull MJ (1998) Deep drawing characteristics of automotive aluminum alloys. Mater Sci Eng A 256:69–82. doi: 10.1016/S0921-5093(98)00845-4

20. Colgan M, Monaghan J (2003) Deep drawing process: analysis and experiment. J Mater Process Technol 132:35–41. doi: 10.1016/S0924-0136(02)00253-4

21. Rabahallah M, Bouvier S, Balan T, Bacroix B (2009) Numerical simulation of sheet metal forming using anisotropic strain-rate potentials. Mater Sci Eng A 517:261–275. doi: 10.1016/j.msea.2009.03.078

22. Barros PD, Neto DM, Alves JL, et al (2015) DD3Imp, 3D Fully Implicit Finite Element Solver: Implementation of Cb2001 Yield Criterion. Rom J Tech Sci − Appl Mech 60:105–136.

23. Yoon JW, Barlat F, Dick RE, Karabin ME (2006) Prediction of six or eight ears in a drawn cup based on a new anisotropic yield function. Int J Plast 22:174–193. doi: 10.1016/j.ijplas.2005.03.013

24. Pottier T, Vacher P, Toussaint F, et al (2012) Out-of-plane Testing Procedure for Inverse Identification Purpose: Application in Sheet Metal Plasticity. Exp Mech 52:951–963. doi: 10.1007/s11340-011-9555-3

25. Kim S-H, Kim S-H, Huh H (2002) Tool design in a multi-stage drawing and ironing process of a rectangular cup with a large aspect ratio using finite element analysis. Int J Mach Tools Manuf 42:863–875. doi: 10.1016/S0890-6955(02)00003-2

26. Yoon J-W, Barlat F, Dick RE, et al (2004) Plane stress yield function for aluminum

alloy sheets—part II: FE formulation and its implementation. Int J Plast 20:495–522. doi: 10.1016/S0749-6419(03)00099-8

27. Thuillier S, Manach PY, Menezes LF (2010) Occurence of strain path changes in a two-stage deep drawing process. J Mater Process Technol 210:226–232. doi: 10.1016/j.jmatprotec.2009.09.004

28. Neto DM, Oliveira MC, Alves JL, Menezes LF (2014) Influence of the plastic anisotropy modelling in the reverse deep drawing process simulation. Mater Des 60:368–379. doi: 10.1016/j.matdes.2014.04.008

29. Danckert J (2001) Ironing of thin walled cans. CIRP Ann - Manuf Technol 50:165–168. doi: 10.1016/S0007-8506(07)62096-4

30. Chandrasekharan S, Palaniswamy H, Jain N, et al (2005) Evaluation of stamping lubricants at various temperature levels using the ironing test. Int J Mach Tools Manuf 45:379–388. doi: 10.1016/j.ijmachtools.2004.09.014

31. Schünemann M, Ahmetoglu MA, Altan T (1996) Prediction of process conditions in drawing and ironing of cans. J Mater Process Technol 59:1–9. doi: 10.1016/0924-0136(96)02280-7

32. Yoon JW, Dick RE, Barlat F (2011) A new analytical theory for earing generated from anisotropic plasticity. Int J Plast 27:1165–1184. doi: 10.1016/j.ijplas.2011.01.002

33. Laurent H, Coër J, Manach PY, et al (2015) Experimental and numerical studies on the warm deep drawing of an Al-Mg alloy. Int J Mech Sci 93:59–72. doi: 10.1016/j.ijmecsci.2015.01.009

34. Manach PY, Coër J, Jégat A, et al (2016) Benchmark 3 - Springback of an Al-Mg alloy in warm forming conditions. J Phys Conf Ser 734:22003. doi: 10.1088/1742-6596/734/2/022003

35. Coër J, Manach PY, Laurent H, et al (2013) Piobert-Lüders plateau and Portevin-Le Chatelier effect in an Al-Mg alloy in simple shear. Mech Res Commun 48:1–7. doi: 10.1016/j.mechrescom.2012.11.008

36. Manach PY, Thuillier S, Yoon JW, et al (2014) Kinematics of Portevin–Le Chatelier bands in simple shear. Int J Plast 58:66–83. doi: 10.1016/j.ijplas.2014.02.005

37. Coër J, Bernard C, Laurent H, et al (2011) The Effect of Temperature on Anisotropy Properties of an Aluminium Alloy. Exp Mech 51:1185–1195. doi: 10.1007/s11340-010-9415-6

38. Develay R (1992) Métaux et alliages , matériaux magnétiques et multimatériaux. Tech l'ingénieur Métaux alliages, matériaux magnétiques multimatériaux 11–16.

39. Demirci Hİ, Esner C, Yasar M (2008) Effect of the blank holder force on drawing of aluminum alloy square cup: Theoretical and experimental investigation. J Mater Process Technol 206:152–160. doi: 10.1016/j.jmatprotec.2007.12.010

40. Hou YK, Li YP, Yu ZQ, et al (2012) Review of Research Progress on Galling in Sheet Metal Forming. Key Eng Mater 501:94–98. doi: 10.4028/www.scientific.net/KEM.501.94

41. Pereira MP, Duncan JL, Yan W, Rolfe BF (2009) Contact pressure evolution at the die radius in sheet metal stamping. J Mater Process Technol 9:3532–3541. doi: 10.1016/j.jmatprotec.2008.08.010

42. Pereira MP, Weiss M, Rolfe BF, Hilditch TB (2013) The effect of the die radius profile accuracy on wear in sheet metal stamping. Int J Mach Tools Manuf 66:44–53. doi: 10.1016/j.ijmachtools.2012.11.001

43. Pereira MP, Yan W, Rolfe BF (2010) Sliding distance , contact pressure and wear in sheet metal stamping. Wear 268:1275–1284. doi: 10.1016/j.wear.2010.01.020

44. Moshksar M., Zamanian A (1997) Optimization of the tool geometry in the deep drawing of aluminium. J Mater Process Technol 72:363–370. doi: 10.1016/S0924-0136(97)00196-9

45. Simões VM, Coër J, Laurent H, et al (2013) Sensitivity analysis of process parameters in the drawing and ironing Processes. Key Eng Mater. doi: 10.4028/www.scientific.net/KEM.554-557.2256

46. Hu P, Liu YQ, Wang JC (2001) Numerical study of the flange earring of deep-drawing sheets with stronger anisotropy. Int J Mech Sci 43:279–296. doi: 10.1016/S0020-7403(99)00119-8

47. Xia ZC, Miller CE, Ren F (2004) Springback Behavior of AA6111-T4 with Split-Ring Test. In: AIP Conf. Proc. AIP, pp 934–939

48. Foecke T, Gnaeupel-Herold T (2006) Robustness of the sheet metal springback cup test. Metall Mater Trans A 37:3503–3510. doi: 10.1007/s11661-006-1045-3

49. Laurent H, Grèze R, Manach PY, Thuillier S (2009) Influence of constitutive model in springback prediction using the split-ring test. Int J Mech Sci 51:233–245. doi: 10.1016/j.ijmecsci.2008.12.010

50. ABAQUS (2008) User's Manuals, Version 6.8.

51. Meinders T, Burchitz IA, Bonte MHA, Lingbeek RA (2008) Numerical product design: Springback prediction, compensation and optimization. Int J Mach Tools Manuf 48:499–514. doi: 10.1016/j.ijmachtools.2007.08.006

52. Padmanabhan R, Oliveira MC, Baptista AJ, et al (2007) Study on the influence of the refinement of a 3-D finite element mesh in springback evaluation of plane-strain channel sections. In: AIP Conf. Proc. pp 847–852

53. Bernard C, Coër J, Laurent H, et al (2016) Influence of Portevin-Le Chatelier Effect on Shear Strain Path Reversal in an Al-Mg Alloy at Room and High

Temperatures. Exp Mech 1–11. doi: 10.1007/s11340-016-0229-z

54. Voce E (1948) The relationship between stress and strain for homogeneous deformations. J. Inst. Met. 74:

55. Voce E (1955) A practical strain-hardening function. Metallurgica 51:219–226.

56. Hockett JE, Sherby OD (1975) Large strain deformation of polycrystalline metals at low homologous temperatures. J Mech Phys Solids 23:87–98. doi: 10.1016/0022-5096(75)90018-6

57. Hill R (1948) A Theory of the Yielding and Plastic Flow of Anisotropic Metals. Proc. R. Soc. London A Math. Phys. Eng. Sci. 193:

58. Felder É (1994) Tribologie de l'emboutissage. Ed Tech Ingénieur 22–26.

59. Magny C (2002) Lois de frottement évolutives destinées à la simulation numérique de l'emboutissage. La Rev Métallurgie 145–156.

60. Knibloe JR, Wagoner RH (1989) Experimental investigation and finite element modeling of hemispherically stretched steel sheet. Metall Trans A 20:1509–1521. doi: 10.1007/BF02665507

61. Fromentin S, Martiny M, Ferron G, et al (2001) Finite element simulations of sheet-metal forming processes for planar-anisotropic materials. Int J Mech Sci 43:1833–1852. doi: 10.1016/S0020-7403(01)00011-X

62. Barlat F, Lege DJ, Brem JC (1991) A six-component yield function for anisotropic materials. Int J Plast 7:693–712. doi: 10.1016/0749-6419(91)90052-Z

63. Morestin F, Boivin M (1996) On the necessity of taking into account the variation in the Young modulus with plastic strain in elastic-plastic software. Nucl Eng Des 162:107–116. doi: 10.1016/0029-5493(95)01123-4

64. Yoshida F, Uemori T, Fujiwara K (2002) Elastic–plastic behavior of steel sheets under in-plane cyclic tension–compression at large strain. Int J Plast 18:633–659.

doi: 10.1016/S0749-6419(01)00049-3

65. Chen Z, Bong HJ, Li D, Wagoner RH (2016) The elastic–plastic transition of metals. Int J Plast 83:178–201. doi: 10.1016/j.ijplas.2016.04.009